\documentclass{PoS}

\title{Further Investigation of Massive Landau-Gauge Propagators 
in the Infrared Limit}

\ShortTitle{Massive Landau-Gauge Propagators at Finite Temperature}

\author{Attilio Cucchieri and \speaker{Tereza Mendes}\\
       IFSC, University of S\~ao Paulo, C.P. 369, CEP 13560-970, 
       S\~ao Carlos SP, Brazil. \\
        E-mail: \email{attilio@ifsc.usp.br, mendes@ifsc.usp.br}}

\abstract{
We investigate how the infrared behavior of electric and 
magnetic gluon propagators in Landau gauge
is affected by temperature. More precisely, we perform large-lattice 
simulations in pure SU(2) gauge theory around the transition 
temperature $T_c$ and study the longitudinal (electric)
and transverse (magnetic) gluon propagators in momentum space,
proposing the calculation of screening masses through an Ansatz 
from the zero-temperature case.
Going from zero to nonzero temperature, we see that the longitudinal
gluon propagator $D_L(p)$ is enhanced, with an apparent plateau value 
in the infrared, while the transverse propagator $D_T(p)$ gets 
progressively more infrared-suppressed, with a clear turnover in 
momentum at all nonzero temperatures considered. 
Our data allow us to associate what was previously seen 
as a peak in the infrared value of $D_L(p)$ at $T_c$ to severe 
finite-size effects along the temperature direction. 
In particular, a temporal lattice extent $N_t\geq 8$ seems to be 
needed to study the electric sector around the transition.
Once these systematic errors are eliminated, the infrared behavior of 
the longitudinal propagator
appears to be rather independent of the temperature below the transition.
Above $T_c$, the infrared value of $D_L(p)$ starts to decrease
monotonically with the temperature.
}

\FullConference{The XXVIII International Symposium on Lattice Field Theory, Lattice2010\\
		June 14-19, 2010\\
		Villasimius, Italy}

\def\spose#1{\hbox to 0pt{#1\hss}}
\def\ltapprox{\mathrel{\spose{\lower 3pt\hbox{$\mathchar"218$}}
 \raise 2.0pt\hbox{$\mathchar"13C$}}}
\def\gtapprox{\mathrel{\spose{\lower 3pt\hbox{$\mathchar"218$}}
 \raise 2.0pt\hbox{$\mathchar"13E$}}}

\begin{document}

\section{Introduction}
 At high temperatures, deconfinement is expected to be felt in 
 the long-distance behavior of correlation functions --- such as 
 the (real-space) longitudinal gluon propagator --- as an exponential 
 falloff with the distance, defining a screening length and conversely 
 a screening mass \cite{Linde:1980ts}.
 Although this predicted behavior has been established at high temperatures
 down to around twice the critical temperature $T_c$ \cite{Cucchieri:2001tw}, 
 it is not clear how a screening mass would show up around $T_c$.
 At the same time, recent studies of zero-temperature Landau-gauge 
 propagators on large lattices have shown a (dynamical) gluon mass at $T=0$ 
 (see e.g.\ \cite{Oliveira:2010xc} for a discussion), in agreement with 
 the so-called massive solution of Schwinger-Dyson equations 
 \cite{Aguilar:2008xm}. One can try to use this knowledge to 
 define temperature-dependent masses
 for the region around the critical temperature. In the following,
 we review briefly the lattice results for the zero-temperature case, 
 discuss the expected behavior for nonzero temperature
 and show our preliminary results for the gluon propagator on large
 lattices for several values of the temperature, drawing
 our conclusions. A more detailed analysis and additional data will 
 be presented elsewhere \cite{inprep}.


\section{Expected Behavior}

 At zero temperature, Landau-gauge gluon and ghost propagators are 
 expected to behave according to the so-called Gribov-Zwanziger 
 scenario, which is based on restricting the gauge configurations 
 to the region delimited by the first Gribov horizon,
 where the smallest nonzero eigenvalue of the Faddeev-Popov matrix 
 goes to zero \cite{Zwanziger:1991gz}. Essentially, the infinite-volume 
 limit favors configurations on the first Gribov horizon and, as a
 result, the ghost propagator should become infrared-enhanced,
 inducing long-range effects in the theory. (In Coulomb gauge, 
 the restriction to the first Gribov region causes the 
 appearance of a confining color-Coulomb potential.) In this
 scenario, formulated for momentum-space propagators,
 the long-range features needed to explain the color-confinement 
 mechanism are thus manifest in the ghost propagator, 
 whereas the momentum-space gluon propagator 
 $D(p)$ is {\em suppressed} in the infrared limit. Such a suppression
 is associated with violation of spectral positivity, which is commonly
 regarded as an indication of gluon confinement. In fact, $D(0)$ is 
 originally expected to be zero, corresponding to maximal violation 
 of spectral positivity. The parametrization of this behavior as a
 propagator with a pair of poles with conjugate complex masses was 
 proposed by Gribov, in connection with his study of gauge copies.
 
 Lattice studies have confirmed the suppression of the gluon 
 propagator in the infrared limit and the enhancement of the ghost 
 propagator at intermediate momenta. However, once the investigated
 lattice sizes were large enough, it became clear that the standard 
 procedure for gauge fixing and simulations is {\em not} compatible 
 with the original scenario in the deep infrared regime. Indeed, the 
 gluon propagator attains a finite value as the momentum is taken to
 zero and the enhancement of the ghost propagator does not persist 
 in this limit. We note the very large lattice sizes employed in 
 order to observe such a behavior, $L \approx 20$ fm and larger.
 The status of these zero-temperature simulations has been recently 
 reviewed in \cite{Cucchieri:2010xr}.
 A good fit of the massive behavior for the gluon propagator is 
 obtained from the Gribov-Stingl form, which generalizes the Gribov
 form described above (see e.g.\ \cite{Cucchieri:2003di}).
 In any case, violation of reflection positivity for the real-space
 gluon propagator is observed for all cases studied 
 \cite{Cucchieri:2004mf}.


As the temperature $T$ is turned on, one expects to observe Debye 
screening of the color charge, signaled by screening masses/lengths 
that can in principle be obtained from the gluon propagator 
\cite{Gross:1980br}.
More specifically, chromoelectric (resp.\ chromomagnetic) screening 
will be related to the longitudinal (resp.\ transverse) gluon 
propagator computed at momenta with null temporal component, 
i.e.\ with $p_0 = 0$ (soft modes).
At high temperatures, we expect the real-space longitudinal
propagator to fall off exponentially at long distances,
defining a (real) electric screening mass,
which can be calculated perturbatively to leading order.
Also, according to the 3d adjoint-Higgs picture for dimensional 
reduction, we expect the transverse propagator to show a confining 
behavior at finite temperature, in association with a nontrivial 
magnetic mass (see e.g.\ \cite{Cucchieri:2001tw}).
The ghost propagator, on the other hand, should not depend on $T$.
We note that these propagators are gauge-dependent quantities,
and the (perturbative) prediction that the propagator poles might 
be gauge-independent must be checked, by considering different gauges.

The above expectations have been checked and confirmed for the gluon
propagator at high $T$ for various gauges
\cite{Cucchieri:2001tw,Heller:1997nqa}.
The behavior of Landau-gauge gluon and ghost propagators around the
critical temperature $T_c$ has been investigated in \cite{Cucchieri:2007ta}.
The study showed a stronger infrared suppression for the transverse 
propagator than for the longitudinal one, confirming the
dimensional-reduction picture also at smaller temperatures.
[We note here that a very recent study \cite{Bornyakov:2010nc}
discusses whether this suppression is consistent with $D_T(0)=0$
and investigates Gribov-copy effects for the propagators.]
It was also found that the ghost propagator is insensitive to the 
temperature, as predicted. 
For the longitudinal gluon propagator, a very interesting
behavior was seen: the data approach a plateau (as a function of the
momentum) in the infrared region and, as a function of temperature,
this plateau shows a sharp peak around the critical temperature. 
The exact behavior
around $T_c$ (e.g.\ whether the peak turns into a divergence at infinite
volume) could not be determined, since relatively small lattices 
were used. All studies mentioned so far are for SU(2) gauge 
theory. 
The momentum-space expressions for the transverse and longitudinal 
gluon propagators $D_T(p)$ and $D_L(p)$ can be found e.g.\ 
in \cite{Cucchieri:2007ta}.

Recently, in \cite{Fischer:2010fx}, further simulations around $T_c$ 
confirmed the above results, and lattice data for the gluon propagator
were used to construct an order parameter for
the chiral/deconfinement transition. More precisely, the authors
use a much finer resolution around $T_c$ and consider the SU(2) and
SU(3) cases. A check of their calculation is done for
the electric screening mass, taken as $D_L(0)^{-1/2}$ and
extracted from the data, where only the $p=0$ raw data point is used.
The considered lattice sizes are still moderate.

Of course, even if an exponential fit to the longitudinal gluon
propagator works at high temperature, it is not obvious that this 
should hold at $T\gtapprox T_c$. 
One should therefore consider more general fits.
At $\,T=0$, the momentum-space propagator is well fitted by a
Gribov-Stingl form (see e.g.\ \cite{Cucchieri:2003di}), allowing 
for complex-conjugate poles
\begin{equation}
D_{L,T}(p) \;=\; 
C\,\frac{1\,+\,d\,p^{2 \eta}}
{(p^2 + a)^2 \,+\, b^2}\,.
\label{GSform}
\end{equation}
This expression corresponds to two poles, at
masses $\,m^2 \;=\; a\,\pm\, i b$, where $m \;=\; m_R \,+\, i m_I$.
The mass $m$ thus depends only on $a$, $b$ and not on the normalization $C$.
The parameter $\eta$ should be 1 if the fitting form also describes
the large momenta region (from our infrared data we get $\eta\neq 1$).
For consistency with the usual definition of electric screening mass,
we expect to observe $\,m_I\to 0\,$ ($\,b\to 0\,$) 
for the longitudinal gluon propagator at high temperature.
Clearly, if the propagator has the above form, then
the screening mass defined by $\,D_L(0)^{-1/2} \,=\, \sqrt{(a^2+b^2)/C}\;\,$
mixes the complex and imaginary masses $\,m_R$ and $m_I\,$ and depends 
on the (a priori arbitrary) normalization $C$.


\section{Results}

We have considered the pure SU(2) case, with a standard Wilson action.
For our runs we employ a cold start, performing a projection on positive 
Polyakov loop configurations. 
Also, gauge fixing is done using stochastic overrelaxation and
the gluon dressing functions are normalized to 1 at 2 GeV. 
We take $\beta$ values in the scaling region
and lattice sizes ranging from $N_s =$ 48 to 192 and from $N_t =$ 2 to 16 
lattice points, respectively along the spatial and along the temporal 
directions.
The temperature is given by $\;T = 1/N_t\, a$.

All our data have been fitted to a Gribov-Stingl behavior, as 
described in the previous section (see Eq.\ \ref{GSform}). These
fits are shown below in all plots, whereas a detailed discussion 
of the associated masses $m_R$, $m_I$ will be presented elsewhere 
\cite{inprep}. We generally find good fits to the Gribov-Stingl 
form (including the full range of momenta), with 
nonzero real and imaginary parts of the pole masses in all cases.
For the transverse propagator $D_T(p)$, the masses $m_R$ and $m_I$ 
are of comparable size. The same holds for $D_L(p)$, but in this case
the relative size of the imaginary mass seems to decrease with increasing 
temperature. We also looked at the real-space propagators,
finding clear violation of reflection positivity for the transverse
propagator at all temperatures. For the longitudinal propagator,
positivity violation is observed unequivocally only at zero temperature 
and for a few cases around the critical region, in association with 
the severe systematic errors discussed below.

Our runs were initially planned under the assumption that a temporal
extent $N_t = 4$ might be sufficient to observe the
infrared behavior of the propagators. (Our goal was, then, to increase
$N_s$ significantly, to check for finite-size effects.) For this value 
of $N_t$, the chosen $\beta$ values:
2.2872, 2.299, 2.313, 2.333, 2.505796
yield temperatures respectively of
0.968, 1.0, 1.04, 1.1, 1.936
times the critical temperature $T_c$.
As is clear from Fig.\ \ref{DLTatTc} below, the assumption that
$N_t = 4$ might be enough is {\em not} verified for the longitudinal 
propagator around the critical temperature, especially in the case 
of larger $N_s$.

\begin{figure}
\centerline{
\includegraphics[height=9.8truecm]{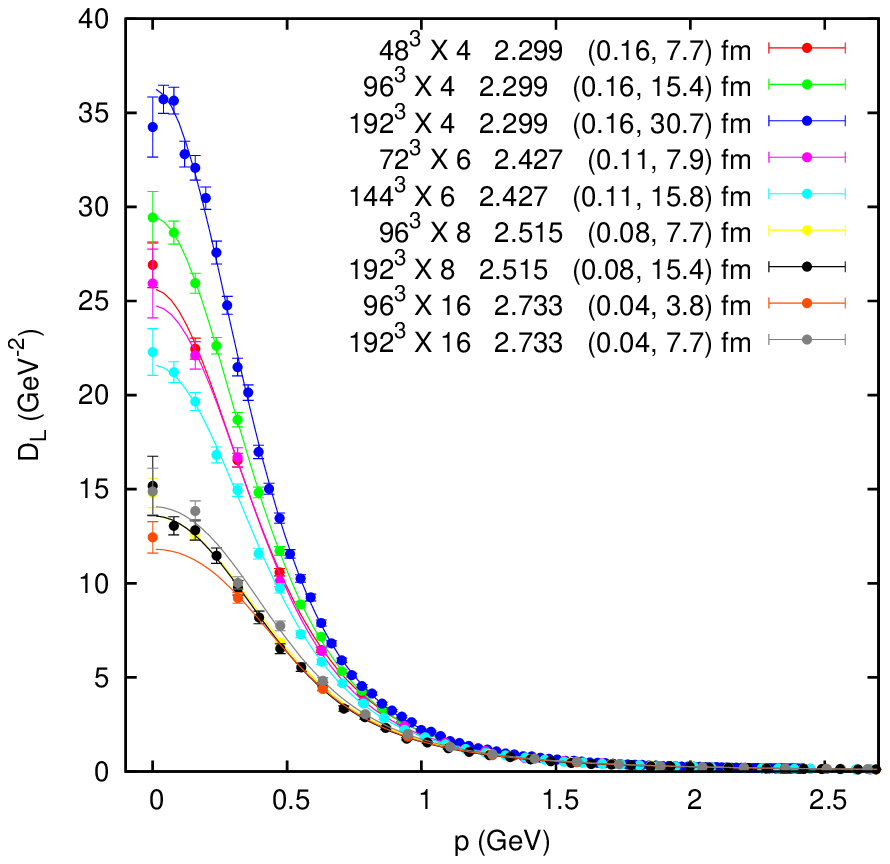}
}
\vskip 4mm
\centerline{
\includegraphics[height=9.8truecm]{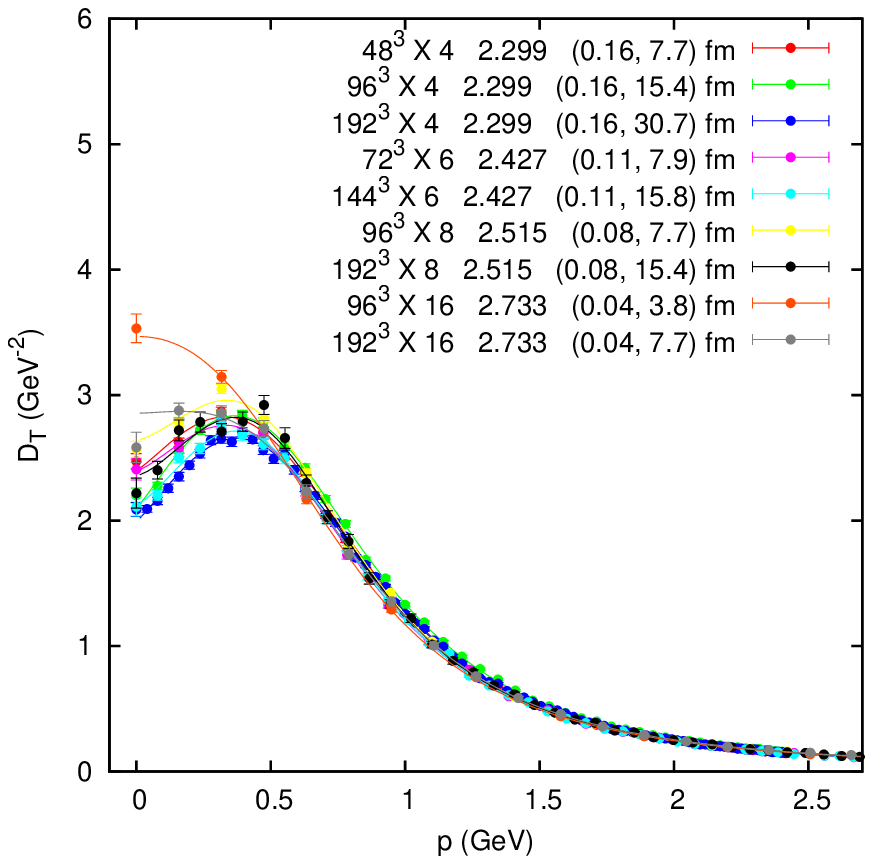}
           }
\caption{Longitudinal (top) and transverse (bottom) gluon propagator at 
$T_c$, for various lattice sizes and values of $\beta$.
Values for $\,N_s^3 \times N_t$, $\,\beta$, lattice spacing $a$ and spatial 
lattice size $L$ (both in fm, in parentheses) are given in the plot 
labels. The resulting temperature is about 302 MeV.}
\label{DLTatTc}
\end{figure}

Indeed, as $N_s$ is doubled from 48 to 96 and then to 192, we see that
the infrared value of $D_L(p)$ changes drastically, resulting in a 
qualitatively different curve at $N_s=192$, apparently with a 
turnover in momentum. Also, in this case the real-space longitudinal 
propagator manifestly violates reflection positivity. We took this 
as an indication that our choice of $N_t=4$ was not valid and therefore 
considered larger values of $N_t$, obtaining a clearer picture of the 
critical behavior of $D_L(p)$. 
As seen in Fig.\ \ref{DLTatTc} (top), once we use large enough
values of $N_t$, the curve stabilizes (within statistical errors) for
four different combinations of parameters. In particular, the two
curves at fixed physical volume (the yellow and the gray curves) agree
very nicely. Note also that the (orange) curve corresponding to the
smallest physical spatial size (i.e.\ 4 fm), may show mild finite-size 
effects. As seen in the bottom plot, the finite-physical-size effects 
are more pronounced for the transverse propagator, which does not seem 
to suffer from the same small-$N_t$ effects. In this case, we see clearly
the strong infrared suppression of the propagator, with a turnover at
around 400 MeV.

\begin{figure}
\hspace*{-1.5cm}
\includegraphics[height=7.2truecm]{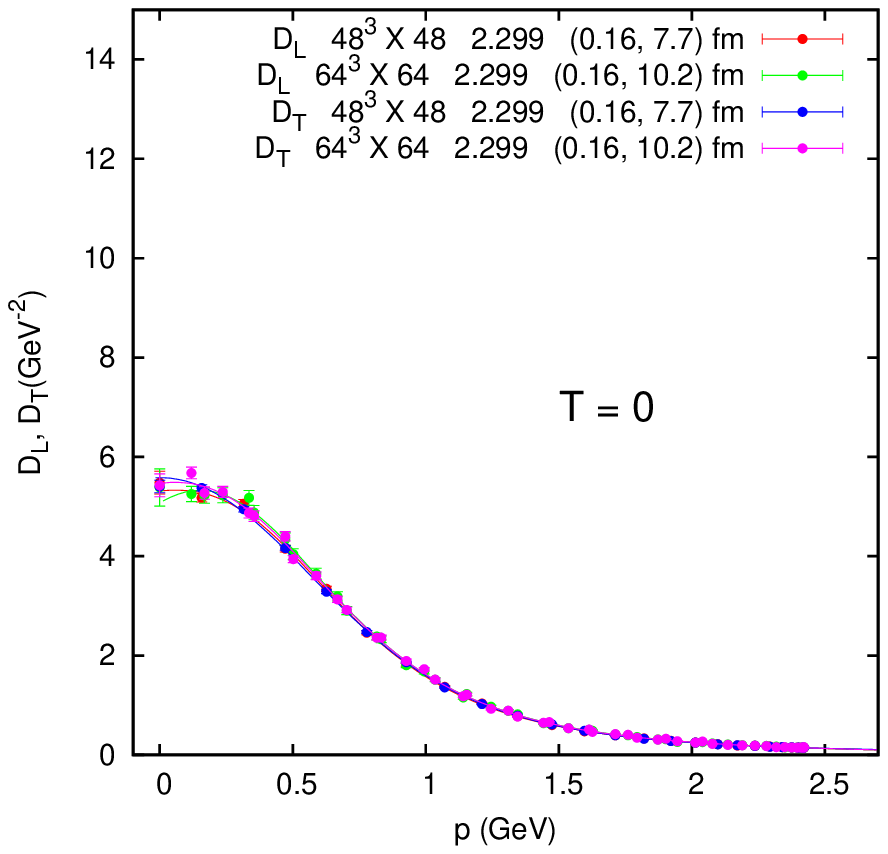}
\hspace*{-2.7cm}
\includegraphics[height=7.2truecm]{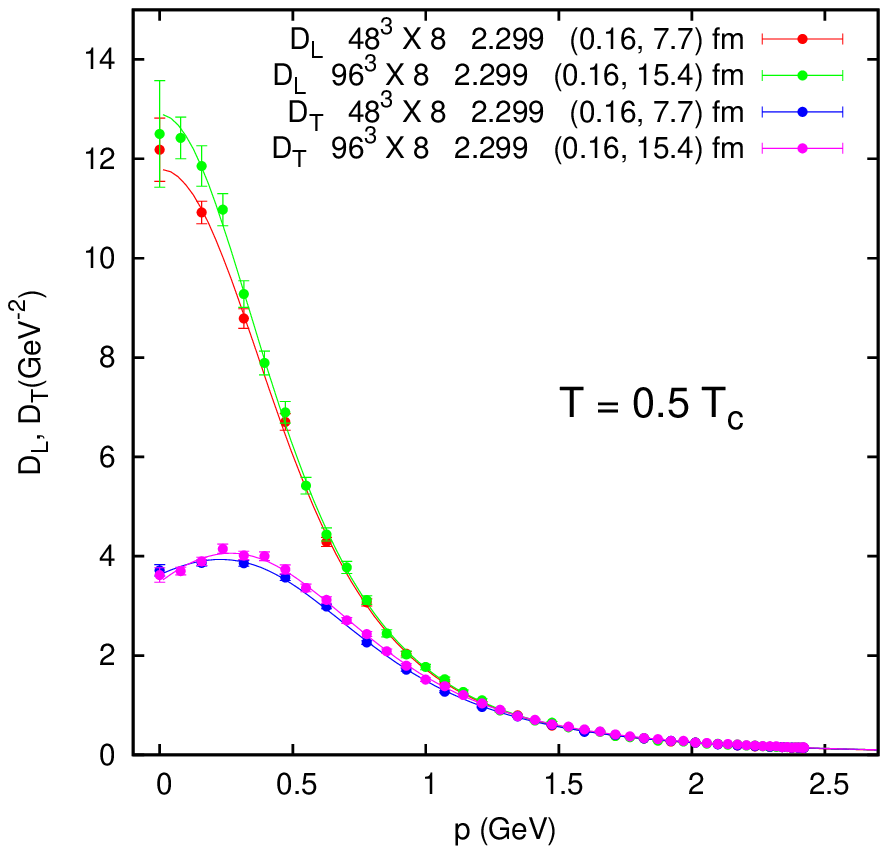}
\caption{Longitudinal and transverse gluon propagators at $T=0$ (left) and
$T=T_c/2$ (right).
Values for $N_s^3 \times N_t$, $\,\beta$, lattice spacing $a$ and spatial 
lattice size $L$ (both in fm, in parentheses) are given in the plot 
labels.}
\label{combined}
\end{figure}

\begin{figure}
\hspace*{-1.5cm}
\includegraphics[height=7.2truecm]{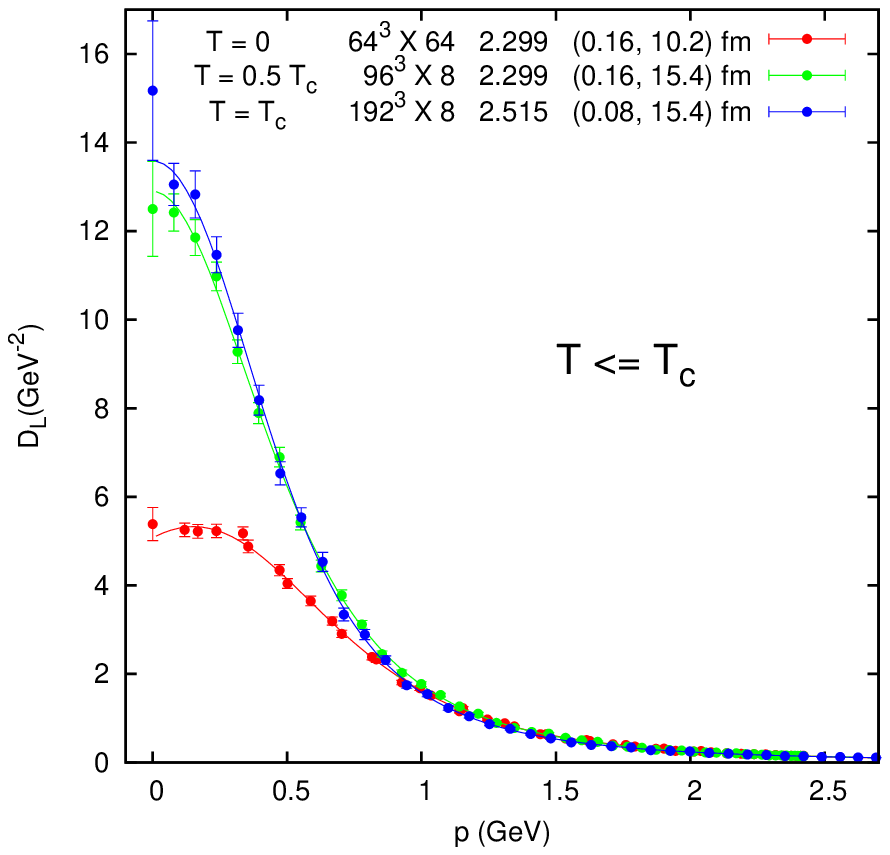}
\hspace*{-2.7cm}
\includegraphics[height=7.2truecm]{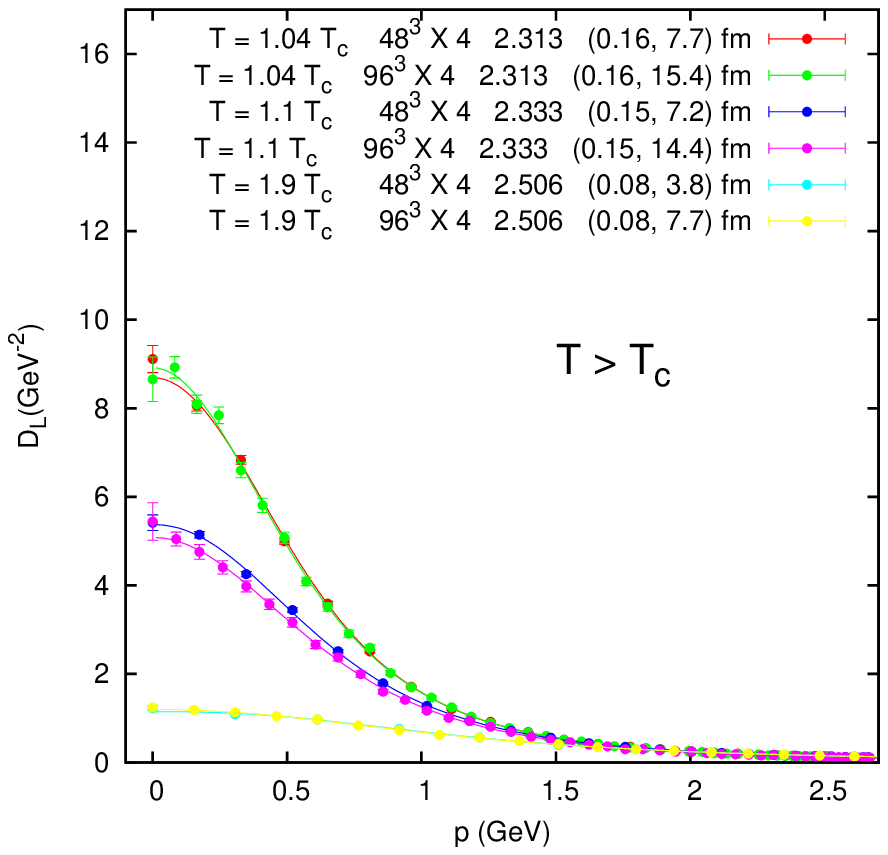}
\caption{Longitudinal gluon propagator
for $T\leq T_c$ (left) and $T>T_c$ (right).
Values for $\,N_s^3 \times N_t$, $\,\beta$, lattice spacing $a$ and spatial 
lattice size $L$ (both in fm, in parentheses) are given in the plot 
labels.}
\label{DLallT}
\end{figure}

The above systematic errors were not observed for other simulated 
values of the temperature (except for the
data at 0.968 $T_c$, not shown here), including the
temperature values just above $T_c$.
We also did runs at $T=0$ and $T\approx T_c/2$, for which we show
combined $D_L(p)$ and $D_T(p)$ data in Fig.\ \ref{combined}.
We see a jump in $D_L(p)$ as temperature is switched on, while 
$D_T(p)$ decreases slightly, showing a clear turnover point at
around 350 MeV.
In Fig.\ \ref{DLallT} we collect data for $D_L(p)$ at several 
temperatures (for $T\leq T_c$, for clarity, we show only valid
lattices with the largest physical size). 
Note that the curve remains unchanged (within errors) from 
$T_c/2$ to $T_c$. Above $T_c$, there is a steady drop.
For all values of $\,T$, $\,D_L(p)$ seems to reach a plateau at small $p$.


\section{Conclusions}

The transverse gluon propagator $D_T(p)$ shows infrared suppression 
and a turnover in momentum (in agreement with the dimensional-reduction 
picture) at all nonzero temperatures considered. The longitudinal 
propagator $D_L(p)$, on the contrary, appears to reach a plateau
at small momenta. The data for $D_L(p)$ are subject to severe 
finite-$N_t$ effects at $\,T\approx T_c$.
As a result, only lattices with $N_t\geq 8$ seem to be free from
systematic errors. After these errors are removed, we see an infrared 
value about 50\% smaller than before. This suggests that there is
no jump in the infrared value of $D_L(p)$ as $T\to T_c\,$ from below 
and that the drop after $T_c$ is significantly smaller than previously 
observed.
[We note that all previous studies of $D_L(p)$ around $T_c$ have 
employed $N_t\leq 4$.] Therefore, the qualitative behavior of $D_L(p)$ 
around the transition has to be revised.

We have obtained good fits of our data to a Gribov-Stingl form, 
with comparable real and imaginary parts of the pole masses, also
in the longitudinal-propagator case. This is in contrast with an
electric screening mass defined by the expression $D_L(0)^{-1/2}$,
which moreover may contain significant finite-size effects. In
that respect, we plan to consider the upper and lower bounds 
for $D(0)$ introduced in \cite{Cucchieri:2007rg}, to investigate the 
infinite-volume limit of the gluon propagator at nonzero temperature.


\section*{Acknowledgements}

The authors thank agencies Fapesp and CNPq for financial support.
Our simulations were performed on the new CPU/GPU cluster at
IFSC--USP (obtained through a Fapesp grant).


\begin{thebibliography}{99}

\bibitem{Linde:1980ts}
  A.~D.~Linde,
  Phys.\ Lett.\  B {\bf 96}, 289 (1980);
  A.~K.~Rebhan,
  Phys.\ Rev.\  D {\bf 48}, 3967 (1993).

\bibitem{Cucchieri:2001tw}
  A.~Cucchieri, F.~Karsch and P.~Petreczky,
  Phys.\ Rev.\  D {\bf 64}, 036001 (2001).

\bibitem{Oliveira:2010xc}
  O.~Oliveira and P.~Bicudo,
  arXiv:1002.4151 [hep-lat].

\bibitem{Aguilar:2008xm}
  A.~C.~Aguilar, D.~Binosi and J.~Papavassiliou,
  Phys.\ Rev.\  D {\bf 78}, 025010 (2008).

\bibitem{inprep}
  A.~Cucchieri and T.~Mendes, to be submitted.

\bibitem{Zwanziger:1991gz}
  D.~Zwanziger,
  Nucl.\ Phys.\  B {\bf 364}, 127 (1991).

\bibitem{Cucchieri:2010xr}
  A.~Cucchieri and T.~Mendes,
  PoS {\bf QCD-TNT09}, 026 (2009).

\bibitem{Cucchieri:2003di}
  A.~Cucchieri, T.~Mendes and A.~R.~Taurines,
  Phys.\ Rev.\  D {\bf 67}, 091502 (2003).

\bibitem{Cucchieri:2004mf}
  A.~Cucchieri, T.~Mendes and A.~R.~Taurines,
  Phys.\ Rev.\  D {\bf 71}, 051902 (2005);
  P.~O.~Bowman {\it et al.},
  Phys.\ Rev.\  D {\bf 76}, 094505 (2007).

\bibitem{Gross:1980br}
  D.~J.~Gross, R.~D.~Pisarski and L.~G.~Yaffe,
  Rev.\ Mod.\ Phys.\  {\bf 53}, 43 (1981).

\bibitem{Heller:1997nqa}
  U.~M.~Heller, F.~Karsch and J.~Rank,
  Phys.\ Rev.\  D {\bf 57}, 1438 (1998);
  A.~Cucchieri, F.~Karsch and P.~Petreczky,
  Phys.\ Lett.\  B {\bf 497}, 80 (2001).

\bibitem{Cucchieri:2007ta}
  A.~Cucchieri, A.~Maas and T.~Mendes,
  Phys.\ Rev.\  D {\bf 75}, 076003 (2007).

\bibitem{Bornyakov:2010nc}
  V.~G.~Bornyakov and V.~K.~Mitrjushkin,
  arXiv:1011.4790 [hep-lat].

\bibitem{Fischer:2010fx}
  C.~S.~Fischer, A.~Maas and J.~A.~Muller,
  Eur.\ Phys.\ J.\  C {\bf 68}, 165 (2010).

\bibitem{Cucchieri:2007rg}
  A.~Cucchieri and T.~Mendes,
  Phys.\ Rev.\ Lett.\  {\bf 100}, 241601 (2008).

\end{thebibliography}
\end{document}